\documentclass[preprint2]{aastex}
\usepackage{natbib}

\usepackage{wrapfig}

\shorttitle{Short-duration Radio Bursts}
\shortauthors{P. Saint-Hilaire, A.O. Benz and C. Monstein}

\begin{document}

\title{Short-duration Radio Bursts with Apparent Extragalactic Dispersion}

\author{P. Saint-Hilaire\altaffilmark{1}, A. O. Benz\altaffilmark{2}, and C. Monstein\altaffilmark{2}}
\affil{Space Sciences Laboratory, University of California, Berkeley, CA 94720, USA}
\affil{Institute for Astronomy, ETH Zurich, 8093 Zurich, Switzerland}

\email{shilaire@ssl.berkeley.edu}

\begin{abstract}
    We present the results of the longest yet undertaken search for apparently extragalactic radio bursts at the Bleien Radio Observatory covering 21000 hours (898 days).
    The data were searched for events of less than 50 ms FWHM duration showing a $\nu^{-2}$ drift in the spectrogram characteristic of the delay of radio waves in plasma.
    We have found five cases suggesting dispersion measures between 350 and 400 cm$^{-3}$ pc while searching in the range of 50 -- 2000 cm$^{-3}$ pc.
    Four of the five events occurred between 10.27 and 11.24 a.m. local civil time. The only exception occurred at night with the full moon in the beam. It was an event that poorly fits plasma dispersion, but had the characteristics of a solar type III burst. However, we were not able to confirm that it was a lunar reflection.
    All events were observed with a log-periodic dipole within 6800 hours but none with  a more directional horn antenna observing the rest of the time.
    These properties suggest a terrestrial origin of the "peryton" type reported before. However, the cause of these events remains ambiguous.
\end{abstract}


\section{Introduction}
Short-duration broadband radio pulses have been predicted from several extragalactic processes, including giant pulses from pulsars \citep{2004ApJ...612..375C},
prompt emission of gamma-ray bursts \citep{1993ApJ...417L..25P} and evaporating mini black holes \citep{1977Natur.266..333R}.
Such impulsive emission is delayed in a plasma relative to the speed of light.
The delay is proportional to the traversed free electron column density, known as the dispersion measure (DM), and to $\nu^{-2}$ where $\nu$ is the observing frequency.
A simultaneously emitted broadband pulse would thus show a characteristic drift in time and frequency, lower frequencies arriving later.

A number of previous searches for dispersed single radio pulses were negative.
Particular galaxies were searched without detection of giant pulses \citep[e.g.][]{2003ApJ...596.1142C}, although  \citet{2013MNRAS.428.2857R} report possible candidates from M31 having a DM of 54.7 pc cm$^{-3}$.
Other searches were aimed at detecting radio pulses associated with gamma-ray bursts.
No radio event was found by \citet{1998A&A...329...61B}, but \citet{2012ApJ...757...38B} report two possible cases delayed from the gamma-ray peak by several minutes.
Several general searches not aimed at particular objects were reported without detections. Amy et al. (1989) operated a transient event monitoring system for a total of 180 days at 843 MHz in parallel to interferometric observations at the Molonglo Observatory.
\citet{2003PASP..115..675K} did not observe any non-solar event within 18 months with the STARE system of three geographically separated instruments at 611 MHz.
\citet{2012ApJ...744..109S} did not detect any extragalactic pulses within 450 hours of a search with the Allen Telescope at 1430 MHz.

Recently four promising detections were reported from a general search with a multi-beam receiver of the Parkes telescope at 1358 MHz \citep{2013Sci...341...53T}.
The events were detected in only one beam of 13, suggesting a non-terrestrial origin.
The range of dispersion measures was from 553 to 1103 pc cm$^{-3}$, the FWHM duration from $<$1.1 to 5.6 ms.
The total observing time of the 13 beams amounted to 23 days each.
In the various Parks searches some 25 other pulses drifting as $\nu^{-2}$ were detected in all beams of the receiver, thus seem to have come into the telescope through sidelobes (Burke-Spolaor et al. 2011).
They are known as "perytons". These events show apparent DMs of 350 - 400 pc cm$^{-3}$, but some deviate from the strict dispersion delay.
The reported fluxes are up to 272 kJy, and the FWHM durations are some 20 ms or more.
An event reported earlier by Lorimer et al. (2007) was observed in several beams, lasted only 4.6 ms, but may be of a similar kind.
Five more events just before a group of Burke-Spolaor events were reported by \citet{2012MNRAS.420..271K} and another four by \citet{2012MNRAS.425.2501B}.
Out of the 25 perytons reported so far, a group of 16 occurred within 30 minutes.
Another two were found within one minute. Thus the number of independent events is 9, all observed at Parkes Observatory in New South Wales, Australia.
Atmospheric conditions such as heavy rain or terrestrial X-ray and gamma-ray events were found not to be associated \citep{2012MNRAS.425.2501B}.
If man-made, these signals traverse the globally protected frequency band at 1420 MHz (21 cm).
Perytons are generally assumed to be terrestrial, but their origin is unknown and they can easily be confused with extragalactic pulses.

Here we report on long-term observations with small antennas having a large field of view but a short range in distance.

\section{Observations and data analysis} \label{sect:selection}

The program ASSERT (Argos Spectrometer Search for Extragalactic Radio Transients) is an observing project executed by the Institute for Astronomy at ETH Zurich. It is an independent system consisting of an antenna, receiver, spectrum analyzer, data taking unit and off-line pulse detection algorithm. The observing time was 24 hours per day.

\subsection{Instrumentation}

The observations were made at the Bleien Observatory of ETH Zurich some 50 km west of Zurich, Switzerland (E 08$^o$ 06' 41'', N 47$^o$ 20' 23''). The receiving antennas that have been used are a commercial log-periodic dipole (R{\&}S HL-040, gain G $\approx$ 4 at 1420 MHz) pointed to the local zenith. At 1500 MHz it has a FWHM beam width of 70$^o$ in East-West and 110$^o$ in North-South. Calibration with the quiet sun yields a conversion factor of 270 kJy K$^{-1}$. The log-periodic antenna was in operation starting on 2009/06/03 until 2010/03/18. It was replaced after this date by a more directive horn antenna having a FWHM beam width of 10$^o$, operating until 2011/11/21. The horn was mounted to the south, scanning the sky at a declination of -14$^o$ 40'.

The receiver is a 16384 channel Fourier transform spectrometer \citep{2005A&A...442..767B} with 8 bit sampling and a bandwidth of 1000 MHz. The effective bandwidth was reduced by filters to avoid terrestrial interference. The useful range was from 1150 to 1740 MHz. The intrinsic frequency resolution is 64 kHz, integrated on-line to 1.02 MHz. One spectrum is obtained every 10 ms.

For both antennas, the system temperature of ASSERT is between 90 and 240 K depending on frequency; the average being 180 K. It was calibrated with the quiet sun and several times per year using an external noise source.

The Bleien Observatory includes also various radio telescopes that continuously observe the radio emission of the Sun from 20 MHz to 850 MHz. These solar dedicated sensitive observations make it easy to identify solar radio bursts detected by ASSERT. The observatory is remotely controlled and has no permanent personnel on site. The activity of persons visiting for service and maintenance is well documented in a log-book. A weather monitoring station records wind, rain, temperature, and humidity. In addition, the low-frequency solar receiving systems ($>$ 200 MHz) are sensitive to pulses caused by terrestrial lightning in all directions and within hundreds of kilometers. Thus atmospheric discharges can be identified over a large frequency range.

\subsection{Data analysis}
Radio bursts of interests were searched using the following methodology:
To ensure that no burst would be cut in half by scanning each file individually,  data chunks (consisting of a central file, with the last half of the preceding file prepended, and the first half of the following file appended) were employed (in effect, all the data was scanned twice). Each data chunk was "cleaned" of interference channels by using standard software developed over the years at ETH Zurich to analyze solar radio data (basically, channels that average out to be
much higher than the average are omitted). In an effort to remove patches of bad pixels, the data was then "de-spiked": each pixel that is more than 25 K (an ad hoc value) above the spectrogram average is checked against its 24 nearest neighbors (a 5x5 window). If no pixel within that 5x5 window is within 10 K (another ad hoc value) of the central pixel, then it is assumed we have a non-continuous source, and is hence discarded. A 1-second gliding average for each channel was then removed to obtain a background-subtracted spectrogram.

The data were then rebinned in 5 ms bins, and were de-dispersed, for each DM between -5000 to +5000 cm$^{-3}$ pc, in 50  cm$^{-3}$ pc increments. In a de-dispersed spectrogram each channel is time-shifted such that a burst with the appropriate DM would appear as a vertical line in the spectrogram. The de-dispersed data are then integrated in time in order to obtain a single time profile.

The DM value for which the time profile's peak is highest is assumed the burst's DM. Such time profiles are shown in Figure 1, as well as their associated spectrograms. The signal-to-noise ratio (S/N) was computed by dividing the peak of the time profile by the standard deviation of the time profile. Likewise, a gaussian was fitted to each peak in the time profiles, yielding a FWHM duration as shown in Table 1. All the burst candidates were ordered by their S/N figure in a master list, from highest to lowest. Candidates with S/N less than 6 were dropped.
This master list comprised of about 22325 candidates. We decided to further restrict ourselves to bursts with less than 50 ms FWHM duration, and with DM greater than 50 cm$^{-3}$ pc, which were examined manually to eliminate the many remaining false positives caused by obviously man-made interference. In the end, only five candidates were found. They are discussed in the next section.

\section{Results} \label{sect:observations}

The five most prominent burst candidates are displayed in Figure 1 and characterized in Table 1. They all have positive DMs. Events with negative DMs may result from superpositions of obvious interference. None of them was continuous through the band contrary to the events shown in Figure 1.

\begin{figure*}[ht!]
        \centering
        \includegraphics[width=15.3cm]{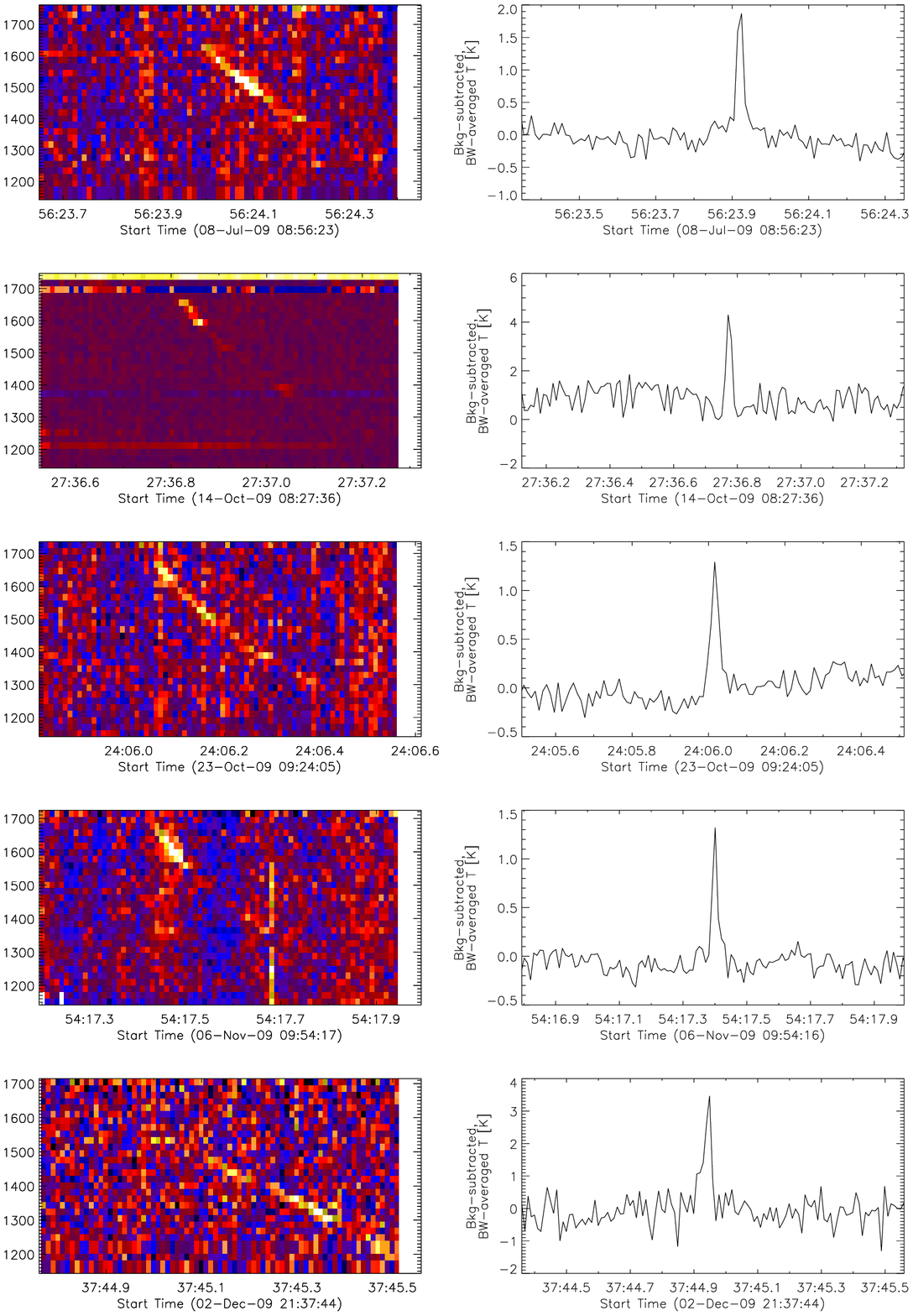}
        \caption{Original spectrograms observed with the ASSERT system and frequency-cumulated de-dispersed lightcurves.
        }
        \label{fig:lcspg}
\end{figure*}

As is obvious from Figure 1, bursts vary strongly in intensity at different frequencies. The peak flux value can easily be an order of magnitude higher than the ``average flux'' values given in Table 1. We note that the five DMs are within a small range between 350 and 400 cm$^{-3}$ pc. They last between 17 and 25 ms; thus they significantly exceed the values reported by \citet{2013Sci...341...53T}.

The last event (Fig.1 bottom) is different in several ways. It appears to be linear in the spectrogram representation, contrary to the first 4 events that are clearly bent. Not surprising as a forward fit to $\nu^{-2}$ is not optimal, the derived duration is longest of all events. The last event occurred at night time, but the other four in a very narrow time interval, which is even narrower in civil time adjusted for daylight saving time. In the latter frame of local civil time, the four events occurred within one hour, between 10.27 and 11.24 a.m.

            \begin{table*}[h!]
            \caption{Detected events in this search, signal-to-noise ratio, dispersion measure DM, and FWHM duration $\Delta t$, and band-averaged flux density}
            \centering
            \begin{tabular}{lllrrrrl}
                \\
                \hline\hline\\
Date &Time&Civil time&S/N&DM&$\Delta t$&Averaged flux&Comment \\
 $[yyyy/mm/dd]$&[UT]&[MEST/MET]&&[cm$^{-3}$pc]&[ms]&[kJy]&\\
\\\hline\\
2009/07/08&08:56:23&10:56:23&10.2&400&21.6&360&\\
2009/10/14&08:27:36&10:27:36&16.4&350&20.2&840&\\
2009/10/23&09:24:06&11:24:06&10.8&350&23.4&250& \\
2009/11/06&09:54:17&10:54:17&10.1&350&17.3&250&\\
2009/12/02&21:37:45&22:37:45&8.3&375&24.9&690&quasi-linear\\
                \\\hline
            \end{tabular}
            \label{tab:events}
            \end{table*}

The following auxiliary data were searched for possible origins:
\begin{itemize}
\item The logbook of the Observatory has no entry for the five days. Thus no personnel was on site and no laboratory equipment in operation.
\item The weather data reporting on rain, humidity, wind and temperature does not show any unusual particularities at the time of the observed events.
\item A possible atmospheric discharge is visible in the solar observations 4.5 minutes before the 2009/07/08 event from 20 - 80 MHz. In the spectrogram of the 2009/11/06 event, a broadband short but not drifting event is present at 17.69 seconds after 09:54:17 UT (Figure 1, fourth from top). However, there was no thunderstorm listed for that day in the local meteorological records.
\item There was no solar event observed by the time of the 4 events that occurred in day time.
\item The 2009/12/02 event occurred at night within 14 hours of full moon. The position of the Moon was within the FWHM beam.
\end{itemize}

\section{Discussion} \label{sect:di}
\citet{2013Sci...341...53T} report a rate of $1.0_{-0.5}^{+0.6}\cdot 10^4$ of their type of bursts per day on the full sky. As the limiting fluence of the very directive Parkes observations, $F_P$, we take their weakest event, 0.6 Jy ms. A small antenna is less sensitive to a point source, but has a larger field of view. The sensitivity limit of ASSERT was tested by adding an artificial event to real background data. An event of 2.5 K antenna temperature lasting 10 ms is detected with a signal-to-noise ratio of 10 for both antennas. The detection limit of ASSERT is thus 25 K ms, or a fluence of 5660 kJy ms for the log-periodic antenna, and 92 kJy ms for the horn antenna. Thus the ASSERT range of detection, $d_A$, is $3.3\cdot 10^{-4}$ times smaller than the Parkes antenna for the log-periodic antenna, and $2.6\cdot 10^{-3}$ times smaller for the horn antenna. On the other hand, ASSERT views a larger angle of the sky.

Assuming a constant rate of events per volume having the same luminosity, the number of observable events, $N(F>F_{min})$, is given by

        \begin{equation} \label{eq:number}
            N(F>F_{min}) = V\ R\  \Delta t\ \ \ ,
        \end{equation}
where $V$ is the volume observed observed by the antenna, $R$ is the rate of events per volume and time, and $\Delta t$ the observing time. $V$ is given by the volume of the spherical sector.

\begin{equation} \label{eq:volume}
            V = {{2\pi d^3 (1-\cos[{\theta\over 2}])}\over 3}\ \ \ ,
        \end{equation}
where $\theta$ is the FWHM beam angle of the antenna. The reported rate of \citet{2013Sci...341...53T} amounts to the events in a sphere with a radius given by the Parks range of detection, $d_P$. Thus
\begin{equation} \label{eq:number_A}
            N_A(F>F_{min}) = \left({d_A^3\over d_P^3}\right)^3 V_P R {1\over 2} (1-\cos({\theta_A\over 2}) \Delta t_A
        \end{equation}
Using the parameters given before, the numbers of expected events during the ASSERT observing times are
\begin{eqnarray} \label{eq:expected}
            N(F>F_{min}) = 4.9\cdot 10^{-5}\ \ \ {\rm log-per\ antenna,}\\
            N(F>F_{min}) = 5.1\cdot 10^{-2} \ \ \ {\rm horn\ antenna.}
        \end{eqnarray}
Figure 2 displays the antenna pointing and half-power beam in equatorial coordinates at the time of the 5 events listed in Table 1. The pointing at the time of the events is widely distributed on the sky. Assuming that the radiation originated from the same source there would be room only within about 10$^o$ from the north celestial pole.

\begin{figure}[ht!]
        \includegraphics[width=7.3cm]{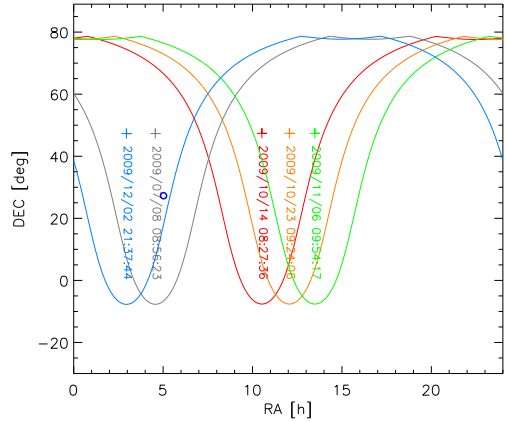}
        \caption{Pointing (plus signs) and half-power beam (curves) of the antenna projected to the sky at the times of the five events reported in Table 1. The position of the full moon on 2009/12/02 21:38 UT is indicated with a circle.
        }
        \label{fig:positions}
\end{figure}

Was the 2009/12/02 event a solar burst reflected by the full moon? Electron beams propagating in the solar corona emit radio waves at the local plasma frequency or its harmonic \citep[review e.g. in][]{2002ASSL..279.....B}. The drift rate of the observing frequency, $\dot\nu$, in a barometric isothermal atmosphere depends linearly on frequency

        \begin{equation} \label{eq:size}
            \dot\nu = - {{\nu v_s \cos \theta }\over{2 H_n (1-\beta \cos \Phi)}} \equiv {\nu\over T}\ \ \ ,
        \end{equation}
where $v_s$ is the beam velocity in upward direction, $\beta = v_s /c$, $H_n$ is the density scale height, $\Phi$ and $\theta$ are the angle between the beam direction and the vertical, resp. the radiation path to the observer. Thus the delay follows the relation
\begin{equation} \label{eq:shift}
            \delta t = {\nu \over T} \delta\nu\ \ \ .
\end{equation}
The drift of a solar electron beam is therefore less bent than the drift due to plasma dispersion,
\begin{equation} \label{eq:shift}
            \delta t = - {{e^2 {\rm DM}} \nu \over {\pi c m_e \nu^3}} \delta\nu\ \ .
\end{equation}

The event 2009/12/02 drifts with a rate -610$\pm$20 MHz s$^{-1}$, where the minus sign indicates a beam propagation in upward direction through the corona. This is in the range of solar Type III bursts in the decimeter region \citep{1983ApJ...271..355B}. Scattering by the Moon reduces the radio flux by a factor of $9.3\cdot 10^{-5}$, assuming a lunar radar cross section of 0.065 \citep{1966JGR....71.4871E}. The observed average flux of the event (Tab. 1) would require a flux of $7.4\cdot 10^9$ Jy before isotropic reflection.

We have searched the RSTN data{\footnote{http://www.ngdc.noaa.gov/stp/space-weather/solar-data/solar-features/solar-radio/rstn-1-second/}} for solar radio emission. However, we were not able to confirm a simultaneous solar burst exceeding $10^4$ Jy at 1415 MHz. We note, however, that on the same day an isolated solar Type III B was observed at the Bleien Radio Observatory at 08:47.6 UT (Solar and Geophysical Data, NOAA) not recorded by RSTN. No flare was reported from GOES X-ray data at the time of both events. For decimetric Type III bursts this is not unusual; on average 61\% are associated with soft X-rays \citep{1985SoPh...97..159A}, but less for isolated bursts.

\section{Conclusions} \label{sect:cl}

Five events of plasma-dispersion like drifts in frequency have been observed in an observation of several years. These events have the following properties:

\begin{itemize}
\item The detection rate is much higher than the one-beam events reported by \citet{2013Sci...341...53T}.
\item The duration is of the order of 20 ms.
\item The dispersion measure is in the range 350 - 400 cm$^{-3}$pc.
\item The bursts vary strongly in intensity at different frequencies.
\item Four out of five bursts occurred between 10.27 and 11.24 a.m. local civil time.
\item One out of five events deviates from the strict dispersion delay.
\item All events were observed with the low-directional log-periodic antenna.
\item No solar events were observed at the time of the events.
\item No laboratory equipment was operated at the time of the events.
\item No lightning or other atmospheric phenomena were recorded on site at the time of the events.
\end{itemize}

The burst characteristics (first 6 points) listed above are consistent with the many-beam events observed with the Parks Telescope, thus showing presumably terrestrial origin \citep{2011ApJ...727...18B}. These authors created the name "perytons" to describe their deceiving similarity with extragalactic events.

Our events were independent and well separated in time. Their rate is about one per 57 days observed with the log-periodic antenna. These events are the first ones of this kind observed outside of the Parks Observatory region. Their origin remains unexplained.

\acknowledgments

We acknowledge the use of RSTN data and thank in particular Kehe Wang and Bon Mills.
Spectrometer and antennas were financially supported by the Swiss National Science Foundation (grants 20-113556 and 200020-121676).

{\it Facilities:} \facility{Bleien Radio Observatory, Gr\"anichen AG, Switzerland}.

\bibliographystyle{apj}

\bibliography{peryton20140203}




\end{document}